\documentclass[oldversion]{aa}  
\usepackage{graphicx}
\usepackage{txfonts}
\begin{document}
   \title{Arp 116: Interacting System or Chance Alignment?}

   \author{Richard de Grijs\thanks{Also Visiting Professor, National
   Astronomical Observatories of China, Chinese Academy of Sciences,
   20A Datun Road, Chaoyang District, Beijing 100012, P.R. China}
          \and
          Alasdair R. I. Robertson
          }

   \offprints{R. de Grijs}

   \institute{Department of Physics \& Astronomy, The University of
   Sheffield, Hicks Building, Hounsfield Road, Sheffield S3 7RH, UK\\
   \email{R.deGrijs@sheffield.ac.uk}}

   \date{Received / accepted}

   \abstract{Using high spatial resolution {\sl Hubble Space
   Telescope}/Advanced Camera for Surveys archival imaging
   observations of \object{Arp 116}, centred on the elliptical galaxy
   \object{NGC 4649}, we explore the novel technique of pixel-by-pixel
   analysis of the galaxy's colour-magnitude diagramme to search for
   any evidence of recent enhanced star formation due to the apparent
   tidal interaction with its spiral companion, \object{NGC 4647}.
   From a detailed analysis of the system's geometry, and based on
   additional circumstantial evidence from extant multi-wavelength
   observations, we conclude that, while there may be grounds for the
   tidal-interaction assumption for this system, any interaction has
   thus far been of insufficient strength to trigger an enhanced level
   of recent star formation in the elliptical component, although
   close inspection of our colour images shows a faint excess of bluer
   pixels (a $\sim 0.20$ mag bluer ``loop'') in the elliptical galaxy
   on the side of the spiral companion. Given that there appears to be
   a moderate reservoir of available gas for ongoing star formation
   (although at low column density), this suggests that we are
   currently witnessing the {\it onset} of the tidal interaction
   between \object{NGC 4647} and \object{NGC 4649}. In addition, the
   triggering of new star formation in \object{NGC 4649} may be
   significantly impeded due to the much lower mass of the spiral
   component.}


   \maketitle
%

\section{Introduction}

One of the major achievements of the {\sl Hubble Space Telescope
(HST)} to date has been the discovery that the progenitors of globular
clusters (GCs), once thought to be the oldest building blocks of
galaxies, continue to form until today (e.g., de Grijs et al. 2003b,
2005). {\sl HST} observations allow us to resolve and study GC systems
far beyond the well-studied GC systems in the \object{Milky Way} and
the \object{Magellanic Clouds}.

Young massive star clusters (YMCs; $M_{\rm cl} \ga 10^5$ M$_\odot$)
have been detected in a great variety of actively star-forming
galaxies. Systems containing a large number of YMCs are most often
interacting spiral-spiral pairs (e.g., the \object{``Antennae''}
galaxies, Whitmore \& Schweizer 1995; \object{NGC 7252}, Whitmore et
al. 1993; the \object{``Mice''} and \object{``Tadpole''} systems, de
Grijs et al. 2003d), since these naturally provide large reservoirs of
gas for star and cluster formation. In fact, massive star {\it
cluster} formation is likely the major mode of star formation in such
extreme starburst environments (cf. de Grijs et al. 2003d). Therefore,
we can use young and intermediate-age massive star clusters as
efficient tracers of the recent violent star formation and interaction
history of galaxies by determining accurate ages for the individual
clusters, even at those ages when morphological interaction features
in their host galaxies have already disappeared.

The formation of YMCs and field stars alike is limited by the supply
of gas. The colliding gas masses associated with interacting galaxies
determine the violence of the star and cluster formation, as well as
the overall gas pressure. Hence the number, and perhaps also the
nature (such as their masses and compactness), of the YMCs, and the
ratio of YMC to field-star formation in particular, is expected to
depend on Hubble type. While observations of gas-rich spiral-spiral
mergers are numerous, star-formation tracers in mixed-pair mergers
have thus far largely been overlooked, although they are undoubtedly
of great interest in the context of the parameter space covered by
the star cluster and field star formation processes.

One of the most important open questions in this field, and one that
we address in this {\it Research Note}, is whether the smaller amount
of gas available for star formation in early-type galaxies might act
as a threshold for star formation in general, and for massive and
compact cluster formation in particular, or possibly result in
enhanced field star formation (including star formation in small
clusters, $M_{\rm cl} \la 10^3$ M$_\odot$) with respect to the mode of
star formation in massive clusters.

The present analysis is of particular current relevance as ever more
sensitive observations increasingly reveal neutral and/or molecular
hydrogen in elliptical and early-type spiral galaxies (e.g., Wiklind
\& Henkel 1989; Cullen et al. 2003, 2006) and in late-stage merger
remnants, which might be the progenitors of present-day elliptical
galaxies (cf. Georgakakis et al. 2001). Star (cluster) formation in
early-type galaxies, long thought to be essentially dust- and gas-free
environments, has therefore returned to the forefront of interest.

\section{Object Selection, Data and Data Reduction}

\subsection{\object{Arp 116}: selection rationale}
\label{selection.sec}

We selected \object{Arp 116}, a combination of the bright ($M_B =
-21.48$ mag) early-type (E2) galaxy \object{NGC 4649} (\object{M60})
and its Sc-type companion \object{NGC 4647} ($M_B = -19.81$ mag), as
our prime target. This is one of the closest systems of mixed-type
interacting galaxies, at a distance of $\sim 16.8$ Mpc in the
\object{Virgo cluster} [Tonry et al. 2001; based on surface brightness
fluctuations (SBF), see Sect. \ref{distances.sec}], and with a
projected separation of $\sim 2\farcm5 \equiv 12$ kpc. At this
distance, the resolution of the Advanced Camera for Surveys (ACS)/Wide
Field Camera (WFC) on board the {\sl HST}, $\sim 0.05$ arcsec
pixel$^{-1}$ (corresponding to $\sim 4$ pc), allows us to (marginally)
resolve and (thanks to their intrinsic brightnesses) robustly identify
individual star clusters of ``typical'' GC size (i.e., with typical
half-mass radii of $R_{\rm hm} \sim 5$ pc).

The \object{Arp 116} system presents an ideal configuration for our
case study into the impact of gravitational interactions on the
triggering of star formation in the early-type member. First, studies
of the nucleus of \object{NGC 4649} by Rocca-Volmerange (1989) show an
excess of far-UV emission, which is indicative of a relatively high
molecular gas density. She postulates this to be either left over from
the main star-formation episode in the galaxy, or the result of
accretion over the galaxy's lifetime. In the context of her ``UV-hot''
model of galaxy evolution, Rocca-Volmerange (1989) interprets the
presence of the far-UV excess in \object{NGC 4649} as caused by a
mostly continuous star-formation rate, although at a level
(independently confirmed by X-ray observations) below her detection
threshold, unless the newly-formed stars are all concentrated in a
very dense cluster of stars. With our new {\sl HST} observations, we
can probe significantly fainter and at higher spatial resolution than
this earlier work, while our passband coverage is also sensitive to
the signatures of recent star formation.

Secondly, Randall et al. (2004) confirm that the X-ray bright
elliptical galaxy is characterised by a dominant excess of bright soft
X-ray emission, predominantly in its nuclear area, at a temperature of
$kT \approx 0.80$ keV (and a model-dependent metal abundance within a
factor of 2 of solar; but see Randall et al. 2006). They interpret
this as thermal emission from interstellar gas, combined with hard
emission from unresolved low-mass X-ray binaries. Moreover,
B\"ohringer et al. (2000) report the galaxy to have an {\it extended}
thermal X-ray halo.

Thirdly, Huchtmeier et al. (1995) and Georgakakis et al. (2001)
determined the total H{\sc i} gas mass in \object{NGC 4649} at,
respectively, $M_{\rm HI} < 1.62 \times 10^8$ and $< 4 \times 10^8$
M$_\odot$. Contamination of this mass by the gas in the late-type
companion is thought to be small. Finally, Cullen et al. (2006), using
CO observations, place an upper limit on the molecular gas mass of
$M_{\rm molec} < 0.72 \times 10^7$ M$_\odot$.

The presence of this moderate amount of atomic and molecular gas,
combined with the likely disturbance caused by the apparent
interaction (possibly linked to enhanced star-formation efficiencies),
suggests that dense gas flows could result in principle, possibly
followed by their turbulent collapse into stars and possibly (massive)
star clusters. In fact, Kundu \& Whitmore (2001; see also Larsen et
al. 2001) used archival {\sl HST}/WFPC2 data of (a section of)
\object{NGC 4649}, in which they detect a significant number (several
tens) of blue clusters -- comprising only the bright end of the GC
luminosity function. Based on their [$0 < (V-I) < 0.5$] mag colours,
these clusters could be as young as 20 Myr for solar metallicity, or
$\la 100$ Myr if their metallicity is as low as 0.02 Z$_\odot$.

\subsection{Observations and reduction}

We obtained archival observations of the NGC 4647/9 system, centred on
\object{NGC 4649}, taken with the ACS/WFC on board {\sl HST} (GO-9401,
PI C\^ot\'e) through the F475W and F850LP broad-band filters on UT
2003 June 17, with exposure times of 750 and 1120 s, respectively.
These filters closely correspond to the Sloan Digitial Sky Survey
(SDSS) $b$ and $i$ bands. The ACCUM imaging mode was used to preserve
dynamic range and to facilitate the removal of cosmic rays.

For the purpose of the research reported on in this {\it Research
Note}, we used the standard on-the-fly data reduction pipeline
(CALACS) in {\sc iraf/stsdas}\footnote{The Image Reduction and
Analysis Facility ({\sc iraf}) is distributed by the National Optical
Astronomy Observatories, which is operated by the Association of
Universities for Research in Astronomy, Inc., under cooperative
agreement with the US National Science Foundation. {\sc stsdas}, the
Space Telescope Science Data Analysis System, contains tasks
complementary to the existing {\sc iraf} tasks. We used Version 3.5
(March 2006) for the data reduction performed in this paper.},
providing us with images corrected for the effects of flat fielding,
shutter shading, cosmic ray rejection and dark current. The use of the
latest flat fields is expected to result in a generic large-scale
photometric uniformity level of the images of $\sim 1$\%. We used the
final, dithered images produced by the most recent release of the {\sl
PyDrizzle} software tool. {\sl PyDrizzle} also performs a geometric
correction on all ACS data, using fourth-order geometric distortion
polynomials, and subsequently combines multiple images into a single
output image and converts the data to units of count rate at the same
time. We registered the individual images obtained for both passbands
to high (subpixel) accuracy, using the Iraf {\sc imalign} routine.

\section{Pixel-by-pixel Analysis}

We constructed colour-magnitude diagrammes (CMDs) on a pixel-by-pixel
basis from our ACS observations to study the field star population.
This technique proved very powerful in our analysis of the ACS early
release observations (ERO) of the \object{``Mice''} and the
\object{``Tadpole''} interacting systems (de Grijs et al. 2003d; see
also Eskridge et al. 2003). Since the ACS/WFC is somewhat undersampled
by its point-spread function at optical wavelengths, the individual
pixels are statistically independent. In our previous work on the ACS
ERO data we identified several subpopulations from the CMDs, which
were found to originate from spatially well-defined regions within the
interacting systems.

We applied the same techniques to the ACS images of the \object{NGC
4647/9} system, in order to search for any evidence of CMD features
corresponding to regions of enhanced star formation in \object{NGC
4649}. Our painstaking analysis of these high-quality {\sl HST}-based
CMD data proved conclusively that \object{NGC 4649} is essentially
composed of a similar stellar population mix throughout the entire
body of the galaxy. The only specific CMD features of note correspond
to (redder) dusty pockets and to the large population of blue and red
GCs (see below).

The galaxy's pixel-CMD does not reveal any significant subpopulation
of pixels that could be ascribed to (an enhanced level of) more recent
star formation, neither throughout the galaxy as a whole nor in
spatially confined regions. For the latter purpose, we carefully
scrutinized the pixel-CMD behaviour of the galaxy in its four
quadrants defined by the major and minor axes, as well as in smaller
wedge-shaped slices, all for galactocentric radii between 20 and 90
arcsec ($\sim 1.6 - 7.2$ kpc). These radial constraints were imposed
on the one hand to avoid the effects of the bright galactic nucleus,
and on the other to avoid spurious pixel-CMD values originating from
the spiral companion, \object{NGC 4647}.

\begin{figure*}
\centering
\includegraphics[width=13.0cm]{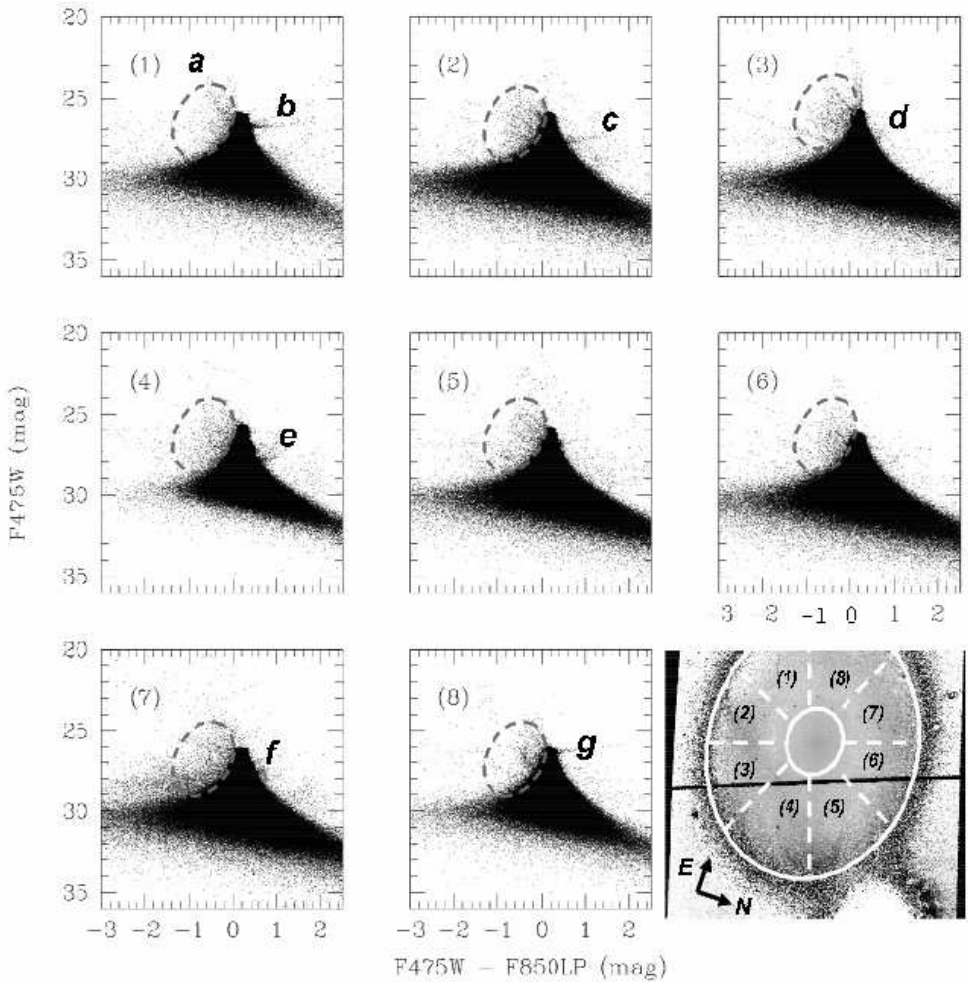}
\caption{Pixel-CMDs of eight wedge-shaped fields in \object{NGC
4649}. The wedge numbers shown in the top left-hand corner of each
panel refer to the wedges indicated on the (F475W -- F850LP) colour
image of the galaxy, included in the bottom right-hand corner of the
figure. The areas of the ``blue excess'' pixels are indicated by the
dashed ellipses in all panels; the linear spurs and features are
identified by labels {\sl a--g}. The colour image shows the full {\sl
HST}/ACS field of view (after geometric corrections), with 4000 ACS
pixels, or 200 arcsec, on a side. The colour levels range linearly
from (F475W -- F850LP) $\simeq -0.6$ (black) to $\simeq 1.3$ (white).}
\label{wedgecmds.fig}
\end{figure*}

In Fig. \ref{wedgecmds.fig} we show the pixel-CMDs for the eight
wedge-shaped areas studied independently in the galaxy. The locations
of which are also indicated here, using a F475W -- F850LP colour image
of the field as guidance (the numbers in each panel correspond to the
numbered wedges superimposed onto the galaxy image). By applying
edge-fitting techniques based on colour histograms at a range of
magnitudes we conclude that both the blue and the red edges of the
individual wedge pixel-CMDs are identical, within the observational
uncertainties of $\Delta ({\rm F475W} - {\rm F850LP}) \la 0.08$ mag,
to some extent depending on pixel magnitude (in the sense of smaller
uncertainties at brighter magnitudes). For additional comparison
purposes, in Table \ref{wedgecols.tab} we list the mean (F475W --
F850LP) colours and standard deviations of the bright-peak pixels, for
$m_{\rm F475W} \le 27.0$ mag. Since these pixels are unaffected by any
recent star formation, the variation in the mean colour reflects the
intrinsic colour variation in the main body of the galaxy. We remind
the reader that the uncertainties are small because of the large
number of data points these statistical measures are based on; they
were obtained by slightly varying the magnitude cut-off used to obtain
the mean values.

\begin{table}
\caption{(F475W -- F850LP) colours and standard deviations of the
pixels with $m_{\rm F475W} \le 27.0$ mag.}
\label{wedgecols.tab}
\centering
\begin{tabular}{c c c}
\hline\hline
Wedge & Mean         & Standard        \\
      & colour (mag) & deviation (mag) \\
\hline
    1 & $0.194 \pm 0.004$ & $0.142 \pm 0.012$ \\
    2 & $0.188 \pm 0.003$ & $0.098 \pm 0.003$ \\
    3 & $0.205 \pm 0.002$ & $0.089 \pm 0.002$ \\
    4 & $0.217 \pm 0.001$ & $0.088 \pm 0.003$ \\
    5 & $0.205 \pm 0.001$ & $0.122 \pm 0.006$ \\
    6 & $0.184 \pm 0.001$ & $0.121 \pm 0.002$ \\
    7 & $0.181 \pm 0.002$ & $0.166 \pm 0.010$ \\
    8 & $0.194 \pm 0.002$ & $0.100 \pm 0.005$ \\
\hline
\end{tabular}
\end{table}

In all cases, and after having taken into account the reddening effect
introduced by Poissonian noise (shot noise) statistics at the faintest
magnitudes, the mean (F475W -- F850LP) colour was compared to the {\sc
galev} simple stellar population (SSP) models (cf. Anders \&
Fritze--v. Alvensleben 2003), which include model sets calculated for
the full set of {\sl HST} imaging filters. The mean colour is
consistent with old stellar populations, for a range of relevant
metallicities (from 0.02 to 1.0 Z$_\odot$).

In Fig. \ref{wedgecmds.fig} we have also indicated the area where we
observe a ``blue excess'' in all wedges (dashed ellipses), as well as
a number of almost linear spurs and feather-like features on both the
blue or red side of the pixel-CMD of the main galaxy body ({\it
a--g}). In all cases, the ``blue excess'' pixels correspond to a
combination of clearly identifiable blue GCs (which appear as
spatially clumped blue excess pixels) and shot noise at the faintest
isophote covered by our detailed investigation (down to $m_{\rm F475W}
= 28.5$ mag); the latter could also be clearly identified based on its
radially random spatial distribution and random noise
characteristics. [In wedge 7, the apparently larger amount of these
blue excess pixels is due to the presence of a foreground star with
similar colours.] We note that there is no clear enhancement of blue
excess pixels on the side of the galaxy facing its spiral companion;
the amount of such pixels is similar or even slightly reduced with
respect to the rest of the galaxy. 

Similarly, the spurs and feathers ({\sl a--g}) correspond to the large
(red and blue) GC population, as well as to shot noise at the faintest
isophote covered by our study. This also applies to the spurs
extending to brighter magnitudes starting from the cut-off magnitude
of the main body of the galaxy in wedges 3 and 4. Once again, the
nature of these pixels which led us to identify them as GCs or noise
is very clear from their spatial distribution and brightness
characteristics. We note that these spurs seem to occur predominantly
on the side of the galaxy away from its spiral companion, which is
consistent with the distribution of the blue excess pixels.

Thus, we conclude that the apparent tidal interaction with \object{NGC
4747} (but see Sect. \ref{verdict.sec} for a discussion on the
robustness of the tidal-interaction assumption) has had a negligible
effect (if any) on the recent star-formation activity in \object{NGC
4649}. However, close inspection of the (F475W -- F850LP) colour image
in the bottom right-hand panel {\it does} show a darker (bluer) loop
of pixels close to the spiral companion. This may be the first
evidence of star formation in the elliptical component induced by the
close encounter between these two galaxies. The mean (F475W -- F850LP)
colour of this blue loop is $-0.04 \pm 0.01$, i.e., approximately 0.20
mag bluer than the pixels in the brightest part of the main body of
the galaxy\footnote{A back-of-the-envelope calculation suggests that
if the colours in Table \ref{wedgecols.tab} represent the old stellar
population, and that (depending on metallicity) a population reddens
by $\sim 1.0$ mag from 10 Myr to 1 Gyr, the blue-loop pixels contain
approximately a $\la 30$ per cent contribution from a young, recently
formed stellar population.}. Since these blue pixels occur at the
faint limit of our observations, they cannot be distinguished in the
individual pixel-CMDs of wedges 4 and 5 because of the overall
increase of the observational uncertainties at low
brightnesses. However, because the blue loop appears to be a distinct
feature in colour space, at a level unexpected from potential
flat-field variations after {\sc MultiDrizzle} and geometric
corrections ($\ll 1$ per cent; Pavlovsky et al. 2006; see also STScI
Analysis Newsletter for ACS, 9 August 2002), we believe this to be a
real feature intrinsic to the galaxy's stellar population. We also
point out that this colour variation of $\sim 0.20$ mag exceeds the
intrinsic colour variations seen in Table \ref{wedgecols.tab}, thus
further suggesting that this is indeed a real feature.

\section{Interaction or chance alignment?}

Our search for young bright blue pixels signifying recent star
formation, particularly in the south-eastern quadrant of the
elliptical, has proven futile. This indicates that there have been no
recent significant episodes of massive star formation within
\object{NGC 4649}, at least not on large spatial scales. This implies
that either the Jeans mass criterion for the formation of massive star
clusters was not met (or, at least, that the threshold for star
formation may not have been met) or that the galaxies may not actually
be interacting.

\subsection{Conditions for the onset of star formation}

There are a number of possible reasons as to why the Jeans mass
criterion, or the threshold conditions for star formation to proceed,
may not have been met in \object{NGC 4649}. The most likely of these
seem to be that (i) there may be insufficient gas, (ii) the gas
density may be too low with respect to the strength of the postulated
gravitational interaction, or (iii) the temperature may be too high
for any collapse to proceed unimpeded. The latter is of concern in
view of the X-ray temperature quoted in Sect. \ref{selection.sec},
although this temperature is comfortably in the range expected for
star-forming and starburst galaxies [see e.g. Hartwell et al. (2004)
for the starburst galaxy \object{NGC 4214}].

Regarding options (i) and (ii), from the example of the large
interacting spiral galaxy \object{NGC 6745a,b} and its much smaller
early-type companion \object{NGC 6745c}, there is evidence that star
and cluster formation in early-type, gas-poor galaxies may be
triggered if the gravitational interaction is sufficiently violent (de
Grijs et al. 2003a). We will return to the issue as to whether the
interaction in the \object{Arp 116} system is sufficiently violent
(and whether the galaxies are sufficiently close to one another) in
Sects. \ref{distances.sec} and \ref{verdict.sec}. However, we point
out here that Cullen et al. (2006) report an upper limit to the
detectable H{\sc i} column density (at the $3\sigma$ level) of $N_{\rm
HI} \la (3.5 \pm 0.3) \times 10^{20}$ cm$^{-2}$. They emphasize that
this is a factor of four lower than that observed in the elliptical
galaxy \object{NGC 1410} by Keel (2004), which does not show strong
evidence of recent star formation either, based on {\sl HST}/STIS
imaging and WIYN spectral mapping techniques. In fact, a conservative
back-of-the-envelope estimate of the mean surface density of the
atomic and molecular gas in \object{NGC 4649}, implies values of $\la
0.1$ M$_\odot$ pc$^{-2}$. We based this estimate on a total gas mass
estimate of $2 \times 10^8$ M$_\odot$ (cf. Sect. 2.1), smoothly
distributed within the galaxy's $D_{25}$ isophote ($7\farcm4 \times
6\farcm0$). This is an order of magnitude lower than the (critical)
threshold densities for star formation derived by Kennicutt (1989) for
the outer regions of spiral galaxies, a situation that is even
worsened if we realise that the H{\sc i} distribution is often
distributed well beyond the optical extent of most galaxies.

This suggests that the possibly weak interaction combined with the low
density of the interstellar gas in the galaxy, may not be sufficiently
conducive to trigger star-formation rates at the level that can be
observed with the current-best instrumentation.

We should also point out that the mass ratio of the \object{NGC
4647/9} system is opposite to that of the \object{NGC 6745} system;
from $K$-band imaging, Cullen et al. (2006) conclude that the ratio of
the masses of the early with respect to the late-type component is
$\sim 5.2$. If a gravitational encounter occurs between unevenly
matched galaxies, provided that they contain sufficient gas
reservoirs, then -- as expected, both intuitively and based on a large
number of dynamical simulations -- the effects of the gravitational
interaction are much more pronounced in the smaller galaxy. For
instance, when we compare the impact of the interaction as evidenced
by star cluster formation (which we take as the most violent mode of
star formation here) between \object{M82} (de Grijs et al. 2001,
2003b,c) and \object{M81} (Chandar et al. 2001), the evidence for
enhanced cluster formation in the larger galaxy is minimal if at all
detectable.

\subsection{The distance to \object{Arp 116}}
\label{distances.sec}

Despite these galaxies being members of the \object{Virgo cluster}
(they are, in fact, projected to be close to the cluster core), their
distance estimates are rather uncertain. Clearly, for the galaxies
to interact, they need to be sufficiently close to one another in
order to respond to their mutual gravitational effects.

Therefore, we embarked on a detailed literature search for reliable
distance measurements to both galaxies. The most comprehensive review
of the distance to \object{NGC 4649}, including an extensive
bibliography up to July 1999, was published by the {\sl HST} Key
Project on the Extragalactic Distance Scale (Ferrarese et
al. 2000a). The secondary distance indicators they used include the
method of SBF, the bright-end cut-off of the Planetary Nebula
Luminosity Function (PNLF), and the peak of the GC Luminosity Function
(GCLF); homogeneous calibration relations of these methods, relating
them to the Cepheid distance scale, were published by Ferrarese et
al. (2000b). Independent analysis of their sample galaxies based on
the $D_n-\sigma$ method resulted in fully consistent distance
measurements (Kelson et al. 2000), thus instilling confidence in the
robustness of these distance calibrations. Here we will discuss the
additional relevant evidence regarding the distance to \object{NGC
4649} presented in studies following the 1999 review.

Ferrarese et al. (2000a,b) adopted as the best average distance
modulus to \object{NGC 4649}, $m-M = 31.09 \pm 0.08$ mag. Subsequent
measurements largely agree with this value (e.g., Tonry et al. 2001;
Di Criscienzo et al. 2006; Mar\'\i n-Franch \& Aparicio 2006; but see
Neilsen \& Tsvetanov 2000). This distance modulus corresponds to a
physical distance to \object{NGC 4649} of $D = 16.52 \pm 0.60$
Mpc. For comparison, if we use the galaxy's systemic velocity, $v_{\rm
sys} = 1117 \pm 6$ km s$^{-1}$ (Ferrarese et al. 2006), and a Hubble
constant H$_0 = 67$ km s$^{-1}$ Mpc$^{-1}$, the corresponding distance
is in close agreement, at $D_{{\rm H}_0} = 16.67 \pm 0.09$ Mpc
(although we point out that large-scale peculiar motions may affect
this result).

In addition, Larsen et al. (2001) used {\sl HST}/WFPC2 data to obtain
the \object{NGC 4649} GCLF based on 345 GCs. They deduced a turn-over
magnitude of $m_V = 23.58 \pm 0.08$ mag for the entire GC sample, and
$m_V = 23.46 \pm 0.13$ mag for the GCs in the blue peak, which are
usually associated with the oldest GC population in a galaxy making up
the ``universal'' GCLF (e.g., Fritze--v. Alvensleben 2004). Using the
Ferrarese et al. (2000b) distance calibration of GCLFs in the
\object{Virgo cluster}, these turn-over magnitudes correspond to
distances of $D_{\rm GCLF} = 17.22^{+2.64}_{-2.29}$ and $D_{\rm GCLF}
= 16.29 \pm 2.50$ Mpc, respectively. The latest study by Forbes et
al. (2004), using GMOS on Gemini-North, detected 2647 GC candidates;
they find a turn-over magnitude of $m_I = 23.17 \pm 0.15$
mag. However, as we are specifically interested in the distance to
\object{NGC 4649}, we are limited to $V$-band measurements for reasons
of observational robustness. For this reason we convert their $I$-band
turnover magnitude to a $V$-band magnitude, using the {\sc galev} SSP
models. We adopt a Salpeter stellar initial mass function covering a
mass range from 0.1 to 70 M$_\odot$ and a metallicity of 0.2
Z$_\odot$. This yields $(V-I) = 1.03$ at an age of 10 Gyr, and hence
we deduce $m_V \simeq 24.2$ mag. Adopting an absolute turn-over
magnitude of $M_V = -7.3$ (Harris et al. 1991), gives a distance of
20.0 Mpc; using Ferrarese et al.'s (2000b) calibration, we derive a
distance of 22.9 Mpc to the galaxy. Considering the manipulation we
had to go through in order to reach this result, the uncertainty on
these $I$-band GCLF distances is $\ga 4$ Mpc.

We base our initial analysis of the distance to the spiral companion,
\object{NGC 4647}, on the compilation of measurements by Solanes et
al. (2002). We also consider the work by Sch\"oniger \& Sofue (1997),
who carefully compared the reliability of using the Tully-Fisher
relation (TFR) based on both H{\sc i} and CO observations (locally
calibrated using Cepheid distances). All of the Solanes et al. (2002)
measurements are TFR based.

Solanes et al.'s (2002) best distance modulus to \object{NGC 4647},
based on homogenisation of the available distance measurements, $m - M
= 31.25 \pm 0.26$ mag, corresponds to a physical distance of $D =
17.78^{+2.26}_{-2.00}$ Mpc. The spread in distance measurements
obtained for this galaxy is in essence within $\sim 2 \sigma$ of this
value, and straddled by the TFR distances of Sch\"oniger \& Sofue
(1997): $D_{\rm CO} = 17.5$ Mpc and $D_{\rm HI} = 22.0$ Mpc. For
comparison, the galaxy's recessional velocity of $v_{\rm sys} = 1419
\pm 63$ km s$^{-1}$ (from the compilation of Falco et al. 1999; but
note that the optical and H{\sc i} 21-cm velocities differ by $\sim
30$ km s$^{-1}$; cf. Huchtmeier \& Richtler 1986) implies a distance
of $D_{{\rm H}_0} \simeq 20.6$ Mpc (or 18.9 Mpc if we base our
estimate on the recessional velocity of $v_{\rm LG} \simeq 1305$ km
s$^{-1}$ with respect to the barycentre of the \object{Local Group};
Helou et al. 1984).

Thus, here we conclude that the two galaxies are most likely within
1--1.5 Mpc of each other. Their velocity differential of order 300 km
s$^{-1}$ (e.g., Gavazzi et al. 1999; Cullen et al. 2006) is also well
within the velocity dispersion of the \object{Virgo cluster} as a
whole, $\sigma_{\rm Virgo} \sim 821$ km s$^{-1}$ (e.g., Sandage \&
Tammann 1976). It is, however, unclear whether the galaxies are
approaching or receding from each other; even very careful modelling
of their luminosity profiles in search of extinction features at large
radii (White et al. 2000) proved inconclusive as to which galaxy might
be in front of the other.

\section{The final verdict?}
\label{verdict.sec}

Owing to the uncertainties in the absolute calibration of the various
secondary distance indicators we needed to rely on in
Sect. \ref{distances.sec}, combined with the relatively low gas
content in the elliptical galaxy, we can neither rule out nor confirm
that both galaxies are in fact sufficiently close to expect signs of
gravitational disturbances\footnote{In fact, we realise that the
\object{Milky Way} and \object{M31} are also well inside the
uncertainty range derived in Sect. \ref{distances.sec}, yet show no
sign of any mutual interaction...}.

Rubin et al. (1999) point out that, although their long-slit optical
spectra show that the kinematics of the spiral component, \object{NGC
4647}, appear nearly normal (but see Young et al. 2006), the H$\alpha$
and molecular (CO and H$_2$) gas disks are clearly asymmetric
(Koopmann et al. 2001; Young et al. 2006), while the more extended
H{\sc i} disk is also slightly extended toward \object{NGC 4649}
(Cayatte et al. 1990), but not nearly by as much as the molecular gas
(Young et al. 2006). This might reflect an H{\sc i} infall scenario,
as Cayatte et al. (1990) suggest, or possibly the {\it onset} of a
tidal interaction between the two galaxies (Rubin et al. 1999; Cullen
et al. 2006). The latter idea is supported by results from
observations as well as numerical simulations that the most apparent
effects of ongoing tidal encounters tend to occur {\it after} the
period of closest approach (e.g., Moore et al. 1998; Rubin et
al. 1999; Young et al. 2006). However, from their detailed analysis of
the gas pressure across the disk of \object{NGC 4647}, Young et
al. (2006) conclude that ram pressure effects alone, postulated to be
caused by tidal forces from \object{NGC 4649}, do not explain the
observations entirely satisfactorily. Instead, they favour a lopsided
gravitational potential as the primary cause for the asymmetries
observed, akin to those commonly seen in spiral galaxies.

Bender et al. (1994) analysed the resolved optical kinematic profiles
of \object{NGC 4649} along both its major and minor axes, out to radii
of $\sim 40$ arcsec. Interestingly, they found evidence for a weak
asymmetry in the line-of-sight velocity dispersion (LOSVD), as well as
negative $h_3$ values (i.e., third-order Gauss-Hermite coefficients,
giving higher-order kinematic information about the deviations of the
LOSVD from a Gaussian distribution), along the galaxy's major axis
(roughly pointing toward the spiral companion). On the other hand, the
minor-axis kinematics do not show any significant asymmetries.
Negative $h_3$ values are normally seen in disky elliptical galaxies,
and in ellipticals with kinematically decoupled cores, but Bender et
al. (1994) did not find any photometric or kinematic evidence for the
presence of such components. In addition, they concluded that the
observed asymmetry was unlikely caused by projection effects. These
conclusions are corroborated by both De Bruyne et al. (2001) and
Pinkney et al. (2003), who traced the galaxy's kinematics out to $\sim
90$ and 40 arcsec, respectively, along its major axis. De Bruyne et
al. (2001), in particular, detected a $\sim 70$ km s$^{-1}$ difference
between the rotation curves on either side of the galactic centre at
their outermost measured radii, just beyond the galaxy's effective
radius. These results provide circumstantial support for the tidal
interaction idea, with \object{NGC 4647} thought to have caused this
minor disturbance of the \object{NGC 4649} kinematics.

Thus, while there may be grounds for the tidal-interaction assumption
in the case of the \object{Arp 116} system, any interaction has thus
far been of insufficient strength to trigger an enhanced level of
recent star formation in the elliptical galaxy component (with the
possible exception of a $\sim 0.20$ mag bluer ``loop'' of pixels in
the elliptical galaxy on the side of its spiral companion). This was
shown conclusively by the null result we obtained from our careful
analysis of the elliptical galaxy's CMD on a pixel-by-pixel
basis. This suggests that we are currently witnessing the onset of the
tidal interaction between \object{NGC 4647} and \object{NGC 4649}. In
addition, the threshold for new star formation in \object{NGC 4649}
may be a further unmet constraint due to the much lower mass of the
spiral component. Detailed numerical ($N$-body) and hydrodynamical
(SPH) modelling, taking into account both the current shape of the
gaseous morphology of \object{NGC 4647} (as well as entertaining the
possibility of the presence of a lopsided gravitational potential),
and the kinematic disturbances seen along the \object{NGC 4649} major
axis, are required to shed light on the future evolution of this
intriguing system.

\begin{acknowledgements}
We acknowledge stimulating discussions with Simon Goodwin and Peter
Anders. RdG acknowledges an International Outgoing Short Visit grant
to the National Astronomical Observatories of China in Beijing from
the Royal Society, and hospitality and support from Profs. X.Q. Na,
and L.C. Deng, Dr.  J. Na, and Z.M. Jin during the final stages of
this project. We are grateful to the referee for suggesting
improvements to make this paper more robust. This paper is based on
archival observations with the NASA/ESA {\sl Hubble Space Telescope},
obtained from the ST-ECF archive facility. We acknowledge the use of
the HyperLeda database (http://leda.univ-lyon1.fr). This research has
also made use of NASA's Astrophysics Data System Abstract Service.
\end{acknowledgements}

\end{document}